\documentclass[prl,twocolumn,letter,preprintnumbers,nofootinbib,superscriptaddress]{revtex4-1}
\usepackage{graphicx}
\usepackage{dcolumn}
\usepackage{bm}
\usepackage{hyperref}
\usepackage[mathlines]{lineno}

\usepackage{multirow}
\usepackage{url}
\usepackage{float}
\usepackage{comment}

\hypersetup{pdfstartview=FitV,colorlinks=true,linkcolor=blue,citecolor=red,filecolor=black,urlcolor=blue}
\newcommand{\met}{$\cancel E_T$}

\def\issue(#1,#2,#3){{\bf #1}, #2 (#3)}

\catcode`\@=11
\def\lsim{\mathrel{\mathpalette\@versim<}}
\def\gsim{\mathrel{\mathpalette\@versim>}}
\def\@versim#1#2{\vcenter{\offinterlineskip
\ialign{$\m@th#1\hfil##\hfil$\crcr#2\crcr\sim\crcr } }}
\catcode`\@=12

\catcode`@=12

\newcommand{\newc}{\newcommand}
\newc{\wt}{\widetilde}
\newc{\ra}{\rightarrow}
\def\beq {\begin{equation}}
\def\eeq {\end{equation}}
\def\bi {\begin{itemize}}
\def\ei {\end{itemize}}
\def\bea {\begin{eqnarray}}
\def\eea {\end{eqnarray}}

\def \met{\rm E{\!\!\!/}_T}

\newcommand{\br}{\begin{eqnarray}}
\newcommand{\er}{\end{eqnarray}}
\newcommand{\be}{\begin{equation}}
\newcommand{\ee}{\end{equation}}

\newcommand{\ch}{\widetilde \chi^\pm}

\def \ch2p {{\wt\chi_2^+}}

\def \ch2m {{\wt\chi_2^-}}

\newc{\dmchi}{\Delta m_{\wt\chi}}

\def \lspone{\wt\chi_1^0}
\def \mlspone{m_{\lspone}}

\usepackage{array}
\newcolumntype{L}[1]{>{\raggedright\let\newline\\\arraybackslash\hspace{0pt}}m{#1}}
\newcolumntype{C}[1]{>{\centering\let\newline\\\arraybackslash\hspace{0pt}}m{#1}}
\newcolumntype{R}[1]{>{\raggedleft\let\newline\\\arraybackslash\hspace{0pt}}m{#1}}

\def\issue(#1,#2,#3){{\bf #1}, #2 (#3)}

\hypersetup{
   colorlinks=true,       
   linkcolor=blue,        
   citecolor=red,         
   filecolor=magenta,      
   }
\begin{document}
%

\title{Is the light neutralino thermal dark matter in the pMSSM ruled out?}

\author{Rahool Kumar Barman}
\email{rahool.barman@okstate.edu}
\affiliation{Department of Physical Sciences, Oklahoma State University, Stillwater, Oklahoma 74078, USA}

\author{Genevi\`eve B\'elanger}
\email{belanger@lapth.cnrs.fr}
\affiliation{LAPTh, Universit\'e Savoie Mont Blanc, CNRS, B.P. 110,  
F-74941 Annecy Cedex, France}

\author{\\Biplob Bhattacherjee}
\email{biplob@iisc.ac.in}
\affiliation{Centre for High Energy Physics, 
Indian Institute of Science, Bangalore 560012, India}

\author{Rohini M. Godbole}
\email{rohini@iisc.ac.in}
\affiliation{Centre for High Energy Physics, 
Indian Institute of Science, Bangalore 560012, India}

\author{Rhitaja Sengupta}
\email{rhitaja@iisc.ac.in}
\affiliation{Centre for High Energy Physics, 
Indian Institute of Science, Bangalore 560012, India}

%
\begin{abstract}
We explore the parameter space of the phenomenological Minimal Supersymmetric Standard Model~(pMSSM) with a light neutralino thermal dark matter~($m_{\lspone} \leq 
m_h/2$) that is consistent with current collider and astrophysical constraints. We consider both positive and negative values of the higgsino mass parameter ($\mu$).
Our investigation shows that the recent experimental results from the LHC as well as from direct detection searches for dark matter by the LUX-ZEPLIN (LZ) collaboration rule out the $Z$-funnel region for the $\mu>0$ scenario. The same results severely restrict the $h$-funnel region for positive $\mu$, however, the allowed points can be probed easily with few more days of data from the LZ experiment. 
In the $\mu<0$ scenario, we find that very light higgsinos in both the $Z$ and $h$ funnels might survive the present constraints from the electroweakino searches at the LHC, and dedicated efforts from experimental collaborations are necessary to make conclusive statements about their present status.
\end{abstract}

\maketitle

The R-parity conserved~(RPC) scenario of the minimal supersymmetric extension of the Standard Model~(MSSM) has been among the most favourable choices for exploring physics beyond the Standard Model~(BSM). The RPC-MSSM scenario alleviates the ``naturalness" problem~\cite{Gildener:1976ih, PhysRevD.20.2619} in the Standard Model~(SM), while also providing a SM-like Higgs boson~($h$) with mass $m_{h} \sim 125~\mathrm{GeV}$ and a stable lightest supersymmetric particle~(LSP), typically the neutralino $\lspone$, which can be a cold dark matter~(DM) candidate. 
The case of the light neutralino $m_{\tilde{\chi}_1^0}\leq m_{h}/2$ is of special interest since it is kinematically feasible for the SM Higgs boson to decay invisibly through $h \to \tilde{\chi}_1^0 \tilde{\chi}_1^0$, thus providing an additional signature for DM in the Higgs sector.
Several studies have explored the prospect of a light neutralino DM in the constrained MSSM~(cMSSM) and the phenomenological MSSM~(pMSSM) considering the various experimental constraints at the time\,\cite{PhysRevD.37.719,Djouadi:1996mj,Belanger:2000tg,Belanger:2001am,Hooper:2002nq,Belanger:2003wb,Dreiner:2009ic,Calibbi:2011ug,Dreiner:2012ex,Ananthanarayan:2013fga,Calibbi:2013poa,Belanger:2013pna,Han:2014nba,Belanger:2015vwa,Hamaguchi:2015rxa,Cao:2015efs,Barman:2017swy,Pozzo:2018anw,GAMBIT:2018gjo,Wang:2020dtb,KumarBarman:2020ylm,VanBeekveld:2021tgn}. 
Collider experiments, like ATLAS and CMS, have made available the latest results from searches of heavy Higgs bosons\,\cite{ATLAS:2020zms}, direct searches of charginos and neutralinos\,\cite{CMS:2020bfa, ATLAS:2021moa, ATLAS:2021yqv, CMS:2022sfi}, as well as the invisible decay of the SM Higgs boson\,\cite{ATLAS:2022yvh}. 
The XENON-1T, PICO-60, PandaX-4T, and LUX-ZEPLIN (LZ) collaborations have also published limits on the DM direct detection~(DD) cross-sections $-$ both spin-dependent~(SD) and spin-independent~(SI)\,\cite{XENON:2018voc,XENON:2019rxp,PICO:2019vsc,PandaX-4T:2021bab,Aalbers:2022fxq,PandaX:2022xas}. 
Among these, the results from the LZ collaboration are the most stringent ones for the SI DD cross-sections\,\cite{Aalbers:2022fxq}. In lieu of these new and improved results, it becomes crucial to revisit the MSSM parameter space containing light neutralino DM, which can also contribute to the invisible decay of the Higgs boson.

In this Letter, we study the current status of the light neutralino DM in the MSSM for both positive and negative values of the higgsino mass parameter $\mu$.
It is worth noting that the supersymmetric explanation for the discrepancy between the experimentally measured value of the $(g-2)_\mu$ and the SM prediction\,\cite{Muong-2:2021ojo} typically requires $\mu > 0$. Due to the prevalent uncertainties in estimating the hadronic contributions in the SM prediction, we prefer an agnostic attitude towards the sign of $\mu$. As the Large Hadron Collider~(LHC) is gearing up for Run-3 and will start collecting data soon, a careful study of the overall status of this scenario is very timely to identify the interesting regions of the parameter space which can be a focal point of the LHC searches at Run-3.

We consider the pMSSM parameter space with ten free parameters defined at the electroweak scale. Our focus is the light neutralino sector with $\mlspone \leq m_h/2$ such that it can contribute to the invisible decay mode of the Higgs boson. The input parameters which capture the physics of the Higgs and electroweakino sectors are: $M_{1}$, the bino mass, $M_{2}$, the wino mass~($M_{1}$ and $M_{2}$ are collectively referred to as the gaugino masses), $\mu$, the higgsino mass, $\tan\beta$, the ratio of the Higgs vacuum expectation values, $M_A$, the pseudoscalar mass, $M_{\tilde{Q}_{3l}}$, $M_{\tilde{t}_{R}}$, $M_{\tilde{b}_{R}}$, the mass of the third generation squarks, $A_{t}$, trilinear coupling of the stop, and $M_{3}$, the mass of the gluino. We perform a random scan 
over ten input parameters for the pMSSM in the following range:
\begin{eqnarray}
30~{\rm GeV} < M_{1} < 100~ {\rm GeV}, ~1~{\rm TeV} < M_{2} < 3~ {\rm TeV}, \nonumber \\ 
100~{\rm GeV} < |\mu| <~2~ {\rm TeV}, ~2 <  \tan{\beta} < 50, \nonumber \\
~100~{\rm GeV} < M_{A} < 5~{\rm TeV}, ~3~{\rm TeV} < M_{\tilde{Q}_{3L}} < 20~{\rm TeV}, \nonumber \\
3~{\rm TeV} < M_{\tilde{t}_{R}} < 20~{\rm TeV}, ~3~{\rm TeV} < M_{\tilde{b}_{R}} < 20~{\rm TeV}, \nonumber \\ 
-20~{\rm TeV} < A_{t} < 20~ {\rm TeV}, ~2~{\rm TeV} < M_{3} < 5~ {\rm TeV}, \nonumber 
\label{eq:scan}
\end{eqnarray}
while we fix
\begin{eqnarray}
M_{\tilde{Q}_{1,2L}} = M_{\tilde{u}_{1,2R}} = M_{\tilde{d}_{1,2R}} = 5\,{\rm TeV}, ~ A_{u/d/c/s/b} = 0, \nonumber \\
M_{\tilde{L}_{1,2,3L}} = M_{\tilde{e}_{1,2,3R}} = 2\,{\rm TeV}, ~ A_{e/\mu/\tau} = 0. \nonumber
\label{eq:fix}
\end{eqnarray}
Since we are interested in light neutralino with $m_{\tilde{\chi}_1^0}\leq m_{h}/2$, it shall dominantly have bino ($\tilde{B}$) component.
Therefore, we scan $M_1$ in the low mass region. The coupling of $Z$ and $h$ bosons to a pair of $\tilde{\chi}_1^0$ also depends on its higgsino ($\wt{H}$) and wino ($\wt{W}$) components. Therefore, in order to circumvent the overabundance of $\tilde{\chi}_1^0$ as the DM candidate, we require it to have some $\wt{H}$ or $\wt{W}$ component. 
We are primarily interested in the higgsino-like next-to-lightest supersymmetric partner~(NLSP) in the present work due to the existing stronger limits on wino-like NLSPs. Hence, $M_2$ is scanned above 1\,TeV, while $|\mu|$ is varied starting from a comparatively lower value of 100\,GeV, considering both positive and negative values of $\mu$.
Previous studies have shown that the DM relic density and direct detection constraints allow only the $Z$ and $h$ funnel regions\,\cite{KumarBarman:2020ylm,VanBeekveld:2021tgn,Carena:2018nlf}. We have performed a dedicated scan where we dynamically tune the $M_1$ parameter to keep $m_{\tilde{\chi}_1^0}$ within $m_{Z}/2 \pm 5~$GeV and $m_{h}/2 \pm 3~$GeV to populate the funnel regions sufficiently. 
Additionally, we extract the pole mass of the top quark, $M_{t}$, randomly from a gaussian distribution with a central value of 173.21\,GeV and a standard deviation of 0.55\,GeV\,\cite{Olive_2014}.
In total, until this point, we scan over a sample of size $2 \times 10^8$ points.

We use \texttt{FeynHiggs\,2.18.1}\,\cite{Heinemeyer:1998yj,Heinemeyer:1998np,Degrassi:2002fi,Frank:2006yh,Hahn:2013ria,Bahl:2016brp,Bahl:2017aev,Bahl:2018qog} to generate the SUSY spectra corresponding to the various sets of input parameters\,\cite{footnote2}, and to calculate the Higgs boson mass, and decays in the Higgs sector. We assume that the lightest CP-even Higgs boson of MSSM is the SM-like Higgs boson, observed by the ATLAS and CMS collaborations, with a combined measured mass $m_h=125.09\pm0.21({\rm stat})\pm0.11({\rm syst})$\,GeV\,\cite{ATLAS:2015yey}. 
We conservatively allow for a theoretical uncertainty of 3\,GeV and consider the mass range $122{\rm~GeV}<m_{h}<128{\rm~GeV}$.
Our scan starts from very low ${\rm tan}\beta$ values where satisfying the observed Higgs boson mass will require high stop masses and $A_t$. In order to ensure that the latter is not so large as to lead to color and charged breaking minima~(CCB)\,\cite{Camargo-Molina:2013sta,Chowdhury:2013dka,Blinov:2013fta}, we require $|X_t|\lesssim\sqrt{6m_{\tilde{t}_1}m_{\tilde{t}_2}}$\,\cite{Chowdhury:2013dka}, where $X_t=A_t-\mu/{\rm tan}\beta$ and $m_{\tilde{t}_1,2}$ represents the masses of the two stops. We observe that the CCB condition has little effect on our parameter space.

\begin{figure}[t!]
    \centering
    \includegraphics[width=0.40\textwidth]{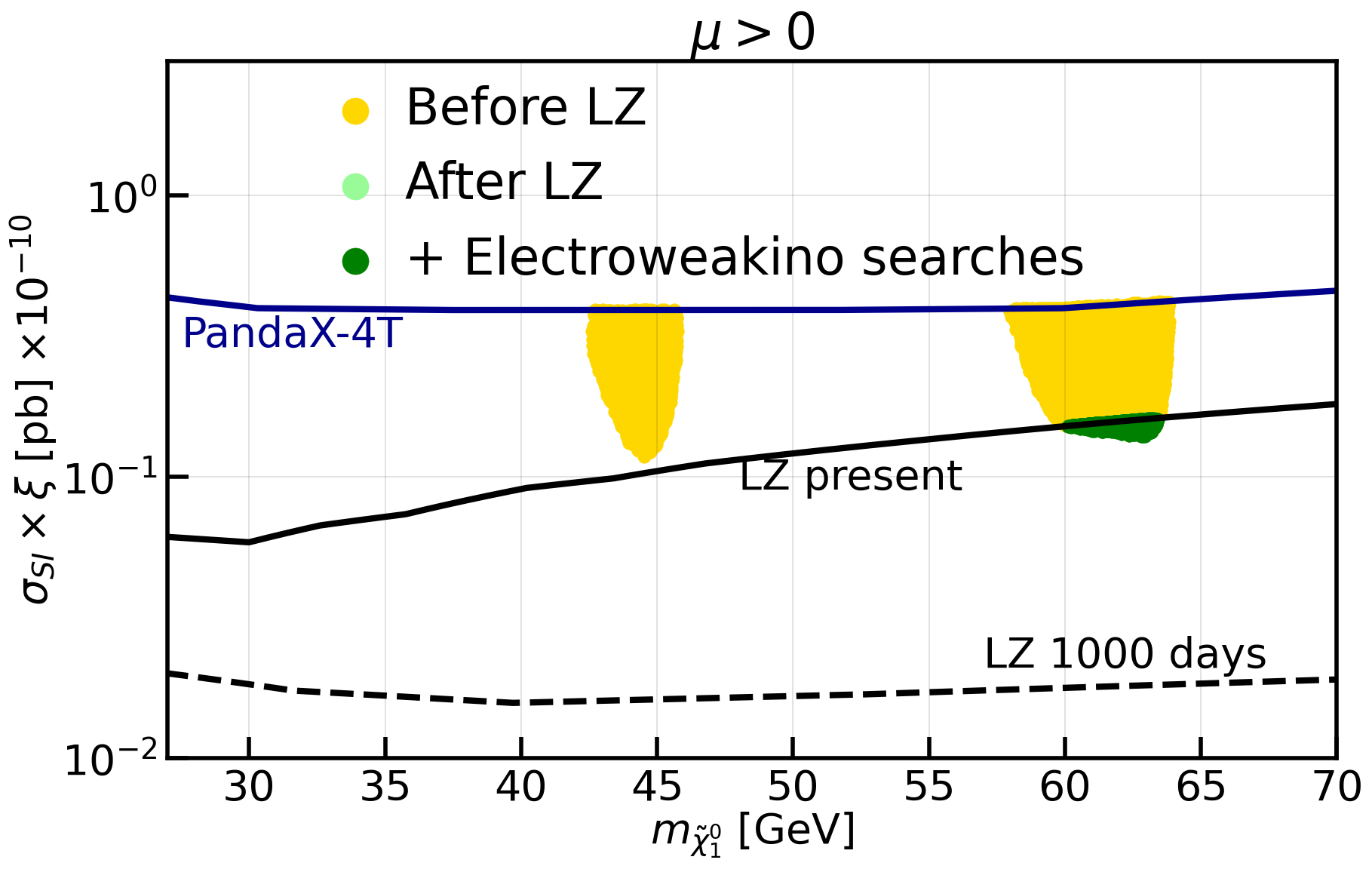}
    \includegraphics[width=0.40\textwidth]{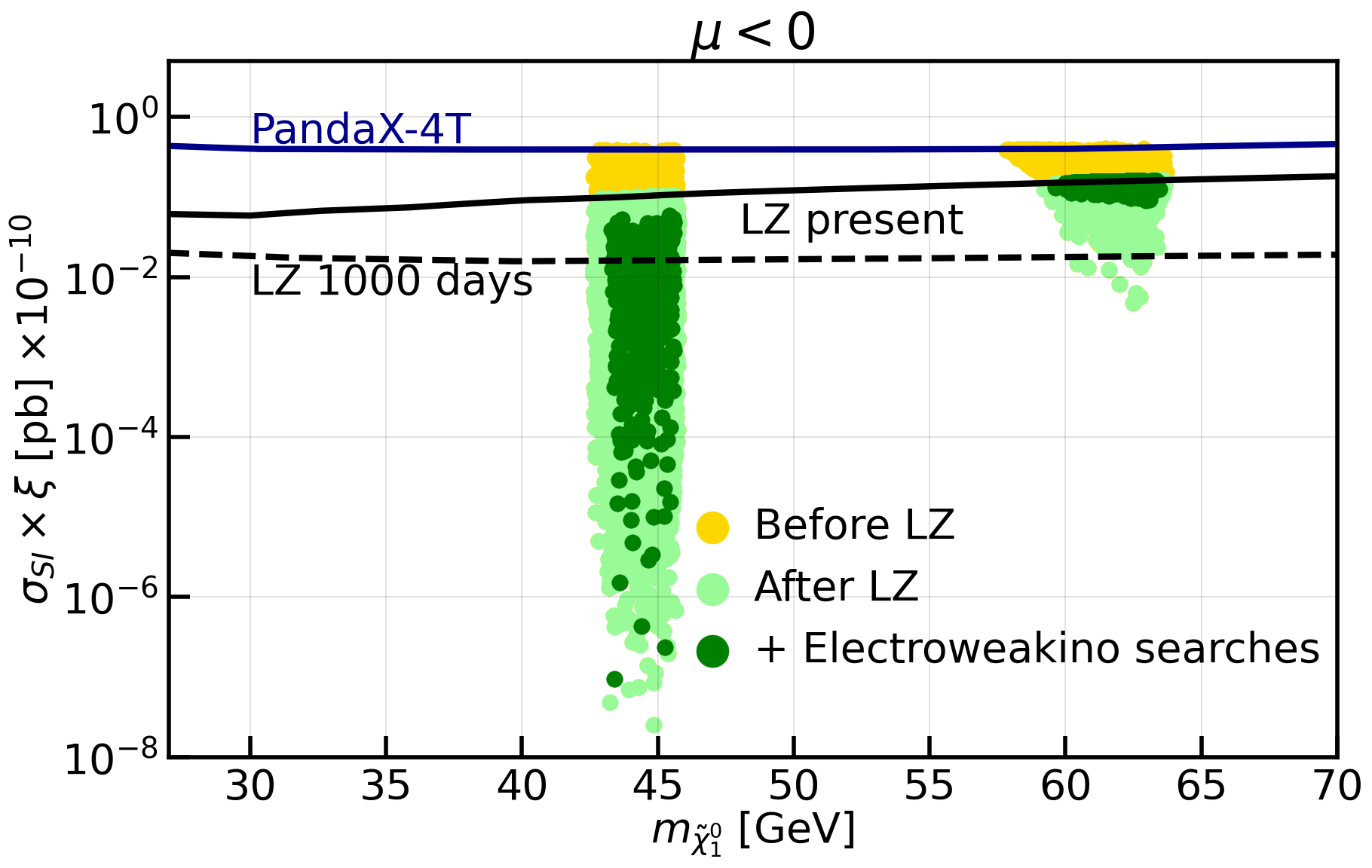}
        \includegraphics[width=0.40\textwidth]{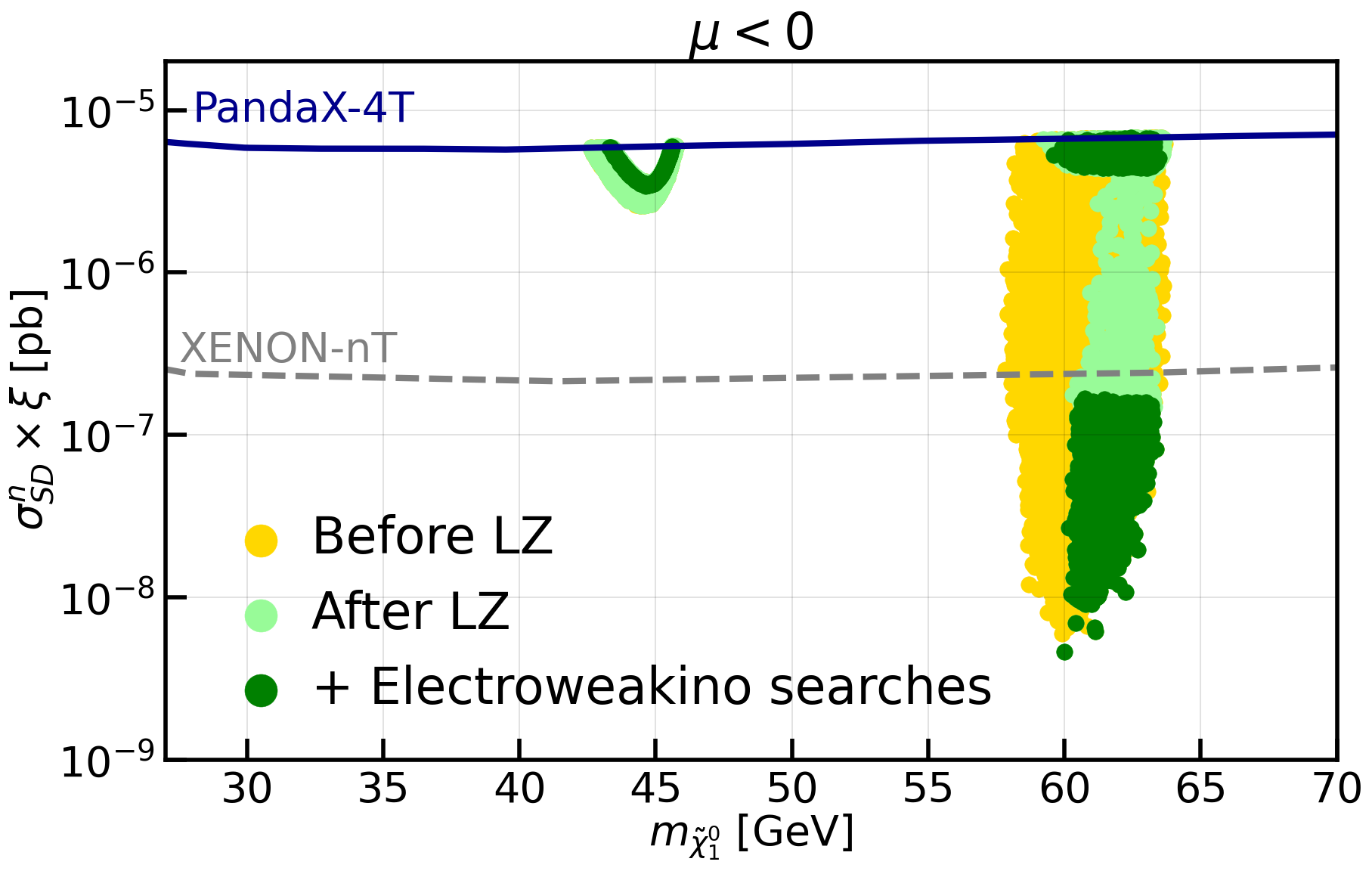}
    \caption{Scaled SI DM-nucleon cross-section ($\sigma_{SI}\times\xi$) for $\mu>0$ ({\it upper}) and $\mu<0$ ({\it center}), along with scaled SD DM-neutron cross-section ($\sigma_{SDn}\times\xi$) for $\mu<0$ ({\it lower}) as a function of the mass of the LSP neutralino DM in the region of parameter space satisfying 
    LEP, flavor, Higgs constraints, relic density, DM DD constraints from the XENON-1T, PICO-60 and PandaX-4T experiments (``Before LZ'' in {\it yellow circles}), ``After LZ'' in {\it light green circles}, overlayed with additional constraints from electroweakino searches at the LHC (in {\it dark green circles}). The current experimental limits from PandaX-4T and LZ are shown ({\it solid lines}) as well as projections for LZ 1000 days and XENON-nT ({\it dashed lines}).}
    \label{fig:DD}
\end{figure}

We apply limits on the partial decay width of the invisible decay of $Z$-boson from new physics, $\Gamma_{\rm inv}^{\rm new}<2$\,MeV\,\cite{ALEPH:2005ab}, chargino mass, $m_{\chi_1^{\pm}}>103$\,GeV\,\cite{OPAL:2003wxm}, and cross-section of associated production of neutralinos in final states with jets, $\sigma(e^+e^-\rightarrow\tilde{\chi}_1^0\tilde{\chi}_2^0)\times{\rm Br}(\tilde{\chi}_2^0\rightarrow\tilde{\chi}_1^0+{\rm jets}) + \sigma(e^+e^-\rightarrow\tilde{\chi}_1^0\tilde{\chi}_3^0)\times{\rm Br}(\tilde{\chi}_3^0\rightarrow\tilde{\chi}_1^0+{\rm jets}) <0.1$\,pb\,\cite{OPAL:2003wxm}, as obtained from experiments at the LEP. 

We also include the flavor physics constraints on various observables, like the branching fractions of processes $b\rightarrow s\gamma$, $B_s\rightarrow \mu^+\mu^-$, and $B\rightarrow \tau\nu$ which are required to satisfy $3.00\times10^{-4}<{\rm Br}(b\rightarrow s\gamma)<3.64\times10^{-4}$\,\cite{HFLAV:2016hnz}, $1.66\times10^{-9}<{\rm Br}(B_s\rightarrow \mu^+\mu^-)<4.34\times10^{-9}$\,\cite{CMS:2014xfa}, and $0.78<({\rm Br}(B\rightarrow \tau\nu))_{\rm obs}/({\rm Br}(B\rightarrow \tau\nu))_{\rm SM}<1.78$\,\cite{Belle:2010xzn}, respectively. We use \texttt{MicrOMEGAS\,5.2.13}\,\cite{Belanger:2004yn,Belanger:2006is,Belanger:2008sj,Belanger:2010gh,Belanger:2013oya,Belanger:2020gnr} to calculate both the LEP and flavor physics observables. 

Additionally, we apply the limits from signal strength measurements of the SM Higgs boson implemented in \texttt{HiggsSignal\,2.6.2}\,\cite{Bechtle:2013xfa, Stal:2013hwa, Bechtle:2014ewa}, as well as limits from heavy Higgs searches at the colliders using the \texttt{HiggsBounds\,5.10.0}\,\cite{Bechtle:2008jh,Bechtle:2011sb,Bechtle:2012lvg,Bechtle:2013wla,Bechtle:2015pma} package. The recent search of heavy Higgs bosons decaying to $\tau$ leptons at ATLAS\,\cite{ATLAS:2020zms} excludes a large part of high ${\rm tan}\beta$ region for $M_A\lesssim1$\,TeV. The parameter space must also satisfy the recent limit on the invisible branching fraction of the SM Higgs boson, Br($h\rightarrow$invisible)$<0.11$\,\cite{ATLAS-CONF-2020-052}. 
We refer to all the constraints related to the Higgs bosons together as the ``Higgs constraints'' hereafter for simplicity, including the SM-like Higgs mass constraint, constraints from \texttt{HiggsSignal\,2.6.2}, \texttt{HiggsBounds\,5.10.0}, and the invisible decay of the SM-like Higgs boson.

The lightest supersymmetric particle~(LSP), $\tilde{\chi}_1^0$, is a viable DM candidate in the MSSM, having a thermal freeze-out production in the early Universe. In the standard cosmology, we require the relic density of the LSP~($\Omega_{\rm LSP}$) to be equal to the observed DM relic density as measured by the PLANCK collaboration $\Omega^{\rm obs}_{\rm DM}h^2 = 0.120\pm0.001$\,\cite{Aghanim:2018eyx}, which assuming a $2\sigma$ interval can vary from 0.118-0.122. Lifting up the requisite that the neutralino LSP forms 100\% of the observed DM relic owing to the possibility of multicomponent DM, we can modify the relic density constraint to $\Omega_{\rm LSP}\lesssim0.122$. \texttt{MicrOMEGAS 5.2.13} is used to compute the relic density of $\tilde{\chi}_1^0$. 

In addition to the relic density constraint, we take into consideration the results from the current DD experiments. These experiments constrain the spin-dependent DM-neutron~(SDn) and DM-proton~(SDp) as well as the spin-independent DD cross-sections of the lightest neutralino LSP ($\tilde{\chi}_1^0$)
as a function of its mass. We use \texttt{MicrOMEGAS 5.2.13} to compute these cross-sections and then compare them with the 90\% confidence level~(CL) upper limits quoted by the XENON-1T (SI\,\cite{XENON:2018voc} and SDn\,\cite{XENON:2019rxp}), PICO-60 (SDp\,\cite{PICO:2019vsc}), PandaX-4T (SI\,\cite{PandaX-4T:2021bab} and SDn\,\cite{PandaX:2022xas}), and LZ (SI,\cite{Aalbers:2022fxq}) experiments. 
The DD limits, typically derived by assuming that a single DM candidate constitutes the entire relic, will weaken in the scenario where the neutralino DM is underabundant by a factor of $\xi = \Omega_{\rm LSP}/0.120$.

Moreover, we must consider the results from direct electroweakino searches at the LHC. We use the \texttt{SModelS\,2.2.1}\,\cite{Kraml:2013mwa,Ambrogi:2017neo,Dutta:2018ioj,Heisig:2018kfq,Ambrogi:2018ujg,Khosa:2020zar,Alguero:2020grj,Alguero:2021dig} package to implement the electroweakino search constraints on our scanned parameter space. 
This version of \texttt{SModelS} includes results from the recent search for electroweakinos in the leptonic final states at CMS\,\cite{CMS:2020bfa} and ATLAS\,\cite{ATLAS:2021moa} and in the hadronic final states at ATLAS\,\cite{ATLAS:2021yqv}, all of which play significant roles in excluding a large range of $m_{\tilde{\chi}_1^{\pm}}$, $m_{\tilde{\chi}_2^0}$ and $m_{\tilde{\chi}_3^0}$, especially with the ATLAS analysis extending the sensitivity to high masses with the hadronic final states.

\begin{figure}[hbt!]
    \centering
    \includegraphics[width=0.40\textwidth]{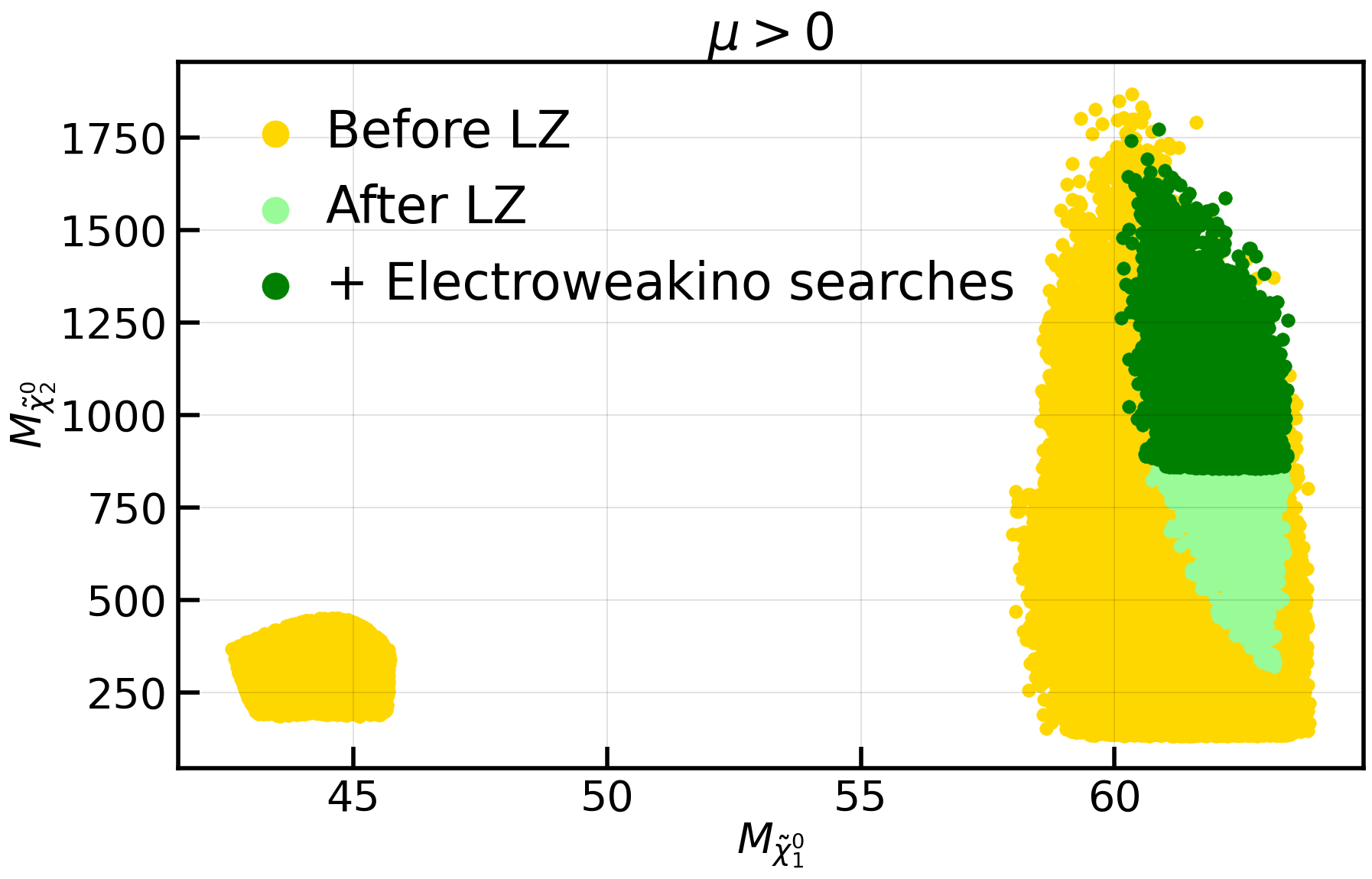}\\
    \includegraphics[width=0.40\textwidth]{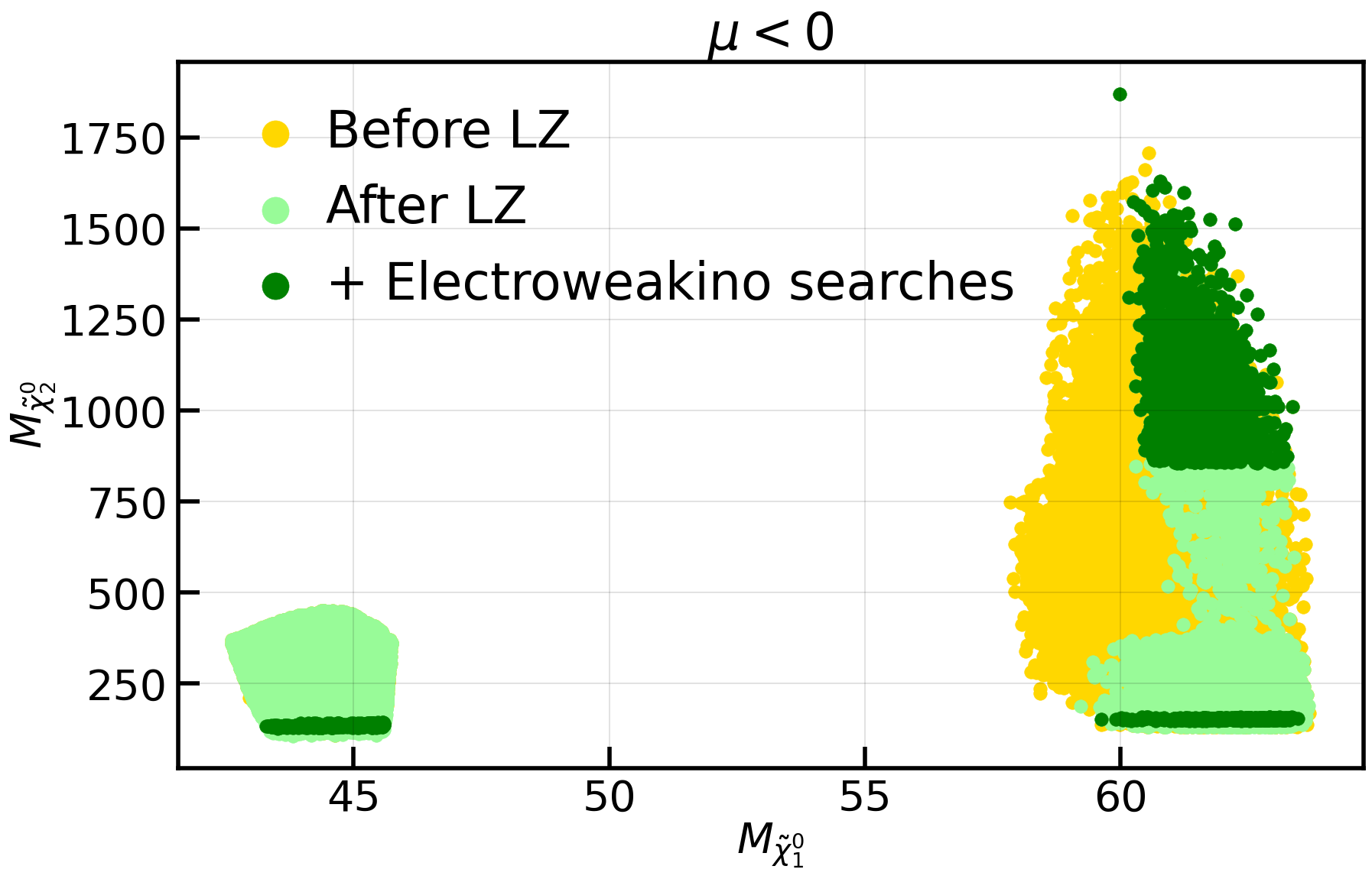}
    \caption{Allowed parameter space for $\mu>0$ ({\it top}) and $\mu<0$ ({\it bottom}) after satisfying the LEP, flavor, Higgs constraints, relic density DM DD constraints from the XENON-1T, PICO-60 and PandaX-4T experiments (``Before LZ'' in {\it yellow circles}), ``After LZ'' in {\it light green circles}, overlayed with additional constraints from electroweakino searches at the LHC (in {\it dark green circles}) in the $m_{\tilde{\chi}_1^0}$-$m_{\tilde{\chi}_2^0}$ plane.}
    \label{fig:mchi12_negmu}
\end{figure}

We apply the constraints on our scanned parameter space in three steps $-$ first \textbf{``Before LZ''} with constraints from LEP, flavor, Higgs constraints, relic density, and the DD experiments XENON-1T, PICO-60, and PandaX-4T, then with the constraint from the LZ experiment, denoted as \textbf{``After LZ''}, and lastly, we add the \textbf{electroweakino constraints} from the LHC.
We identify interesting regions of the parameter space surviving all the constraints and perform dedicated scans over these regions again with an additional sample of size $\sim10^8$, which makes the total size of our scan $3\times10^8$. The {\it upper} panel of Fig.\,\ref{fig:DD} shows the scaled (with $\xi$) SI DD cross-sections for the allowed parameter space with $\mu > 0$ after applying all the constraints from 
``Before LZ'' in {\it yellow}, ``After LZ'' in {\it light green} and from electroweakino constraints in {\it dark green}. We observe that the recent LZ experiment, with only 60 days of data, has played a crucial role in excluding the $Z$-funnel region for positive $\mu$. 
To ascertain the robustness of this result we allow for a 20\% theoretical uncertainty on the relic density. We find that the scaled SI DD cross-section can be reduced below the present LZ limit. However, all such points are excluded by the present electroweakino searches.
In the $h$-funnel we find a region of parameter space surviving all present constraints. In this region the SI-DD cross-section is just below the present limit and is well above the full 1000 days projections, thus the region will be probed with just a few more days of data from the LZ experiment\,\cite{footnote3}. In fact this strong impact of the recent LZ result stands true for any BSM model with a light Majorana fermionic DM having couplings with the $h$ boson in the framework of a simplified model~(see Ref.\,\cite{future-work}). Fig.\,\ref{fig:mchi12_negmu} illustrates the impact of electroweakino searches which restrict $M_{\tilde{\chi}_2^0}\gtrsim850$\,GeV.

Let us now investigate the $\mu < 0$ scenario.
The {\it center} and {\it lower} panels of Fig.\,\ref{fig:DD} respectively illustrate the scaled SI and SDn DD cross-sections for the allowed parameter space with $\mu < 0$, where the {\it colors} have the same meaning as described for Fig.\,\ref{fig:DD}~{\it upper} panel. 
We observe that the recent LZ result and electroweakino searches have excluded most of the $h$-funnel region leaving only a marginally allowed region where the $M_{\tilde{\chi}_2^0}$ is either very small ($\sim$\,140-155\,GeV) or larger than 850\,GeV, as seen from the {\it bottom} panel of Fig.\,\ref{fig:mchi12_negmu}.
We expect this region to be probed in the near future by the LZ experiment with its full 1000-day exposure. On the other hand, the $Z$-funnel is not affected much by the LZ limits and a large fraction lies well below the future reach of LZ. However the entire Z-funnel region is well within the projected reach on $\sigma_{SDn}$ of XENON-nT ({\it bottom panel} of Fig.\,\ref{fig:DD}). In fact XENON-nT will probe the parameter space with light $\tilde{\chi}_2^0$ in both the $Z$ and $h$ funnels. Indeed the SD DD cross-section is proportional to the square of the coupling of the LSP with the Z boson which depends on the higgsino component of $\tilde{\chi}_1^0$. Therefore the lighter higgsinos have higher values of the SDn DD cross-section than the heavier ones, which creates the two separate {\it dark green} patches in the $h$-funnel of the {\it lower} plot of Fig.\,\ref{fig:DD}. 

Fig.\,\ref{fig:mchi12_negmu} reveals a major difference between the $\mu>0$ and the $\mu<0$ scenarios. 
The LZ limit excludes lighter higgsinos ($\lesssim 300$\,GeV) for $\mu>0$, while they survive when $\mu<0$. 
This happens due to a 
cancellation\,\cite{footnote4} between the contributions of the two CP-even neutral Higgs bosons ($h$ and $H$) involving diagrams with down-type quarks to the SI DD cross-section when $\mu<0$ in the MSSM.
On the other hand these two contributions interfere constructively for $\mu>0$, increasing the SI DD cross-sections.
Among the light higgsinos allowed by LZ, the electroweakino searches allow points in a very narrow region of parameter space in both the $Z$ and $h$-funnels for $\mu < 0 $, many of which have very small $R$-values\,\cite{footnote}, as shown in Fig.\,\ref{fig:analysis} of Supplemental Material.
It is interesting to note that if future DD experiments discover a light DM in the $Z$-funnel, it will mostly indicate a negative value of $\mu$.
An observation of DM signal in the $h$-funnel from DD experiments would require an additional observation of a signal in the collider experiments to shed light on the sign of $\mu$ $-$ observation of light higgsinos will hint towards negative $\mu$, however, heavy higgsinos will not be able to lift this ambiguity. 
Moreover, we find that the allowed points in each of these regions are clustered around specific tan$\beta$ ranges $-$ low tan beta values of 3-10 in the $h$-funnel for $\mu>0$ while for $\mu<0$, tan$\beta \sim$ 3-18 in the $Z$-funnel, $\sim$ 3-6 for heavy higgsinos in the $h$-funnel, and $\sim$ 16-50 for light higgsinos in the $h$-funnel.

Representative benchmarks from each of the allowed regions of the parameter space are presented in the Supplemental Material. These benchmarks have very small uncertainty in the Higgs boson mass ($\lesssim \mathcal{O}(1)~$GeV), and have \texttt{SModelS} $R$-values below 0.5. They are also allowed when tested with \texttt{CheckMATE 2}\,\cite{Dercks:2016npn}, another package that implements the constraints from electroweakino searches. We find that the Tevatron limits for light charginos\,\cite{Mario_P_Giordani_2006} are also not sensitive to these benchmarks.
To estimate the prospects for probing the region with light charginos and neutralinos at the LHC, 
we perform an analysis of the low mass higgsino-like electroweakinos in the leptonic $3l+\met$ final state at $\sqrt{s}=14$\,TeV using the \texttt{XGBOOST}\,\cite{xgboost}
framework. 
We study the process $pp\rightarrow\tilde{\chi}_1^{\pm}\tilde{\chi}_2^0/\tilde{\chi}_1^{\pm}\tilde{\chi}_3^0,~\tilde{\chi}_1^{\pm}\rightarrow W^{\pm}\tilde{\chi}_1^0,~\tilde{\chi}_2^0/\tilde{\chi}_3^0\rightarrow f\bar{f}\tilde{\chi}_1^0$ with $m_{\tilde{\chi}_1^{\pm}}=125.1$\,GeV, $m_{\tilde{\chi}_2^0}=129.9$\,GeV, $m_{\tilde{\chi}_3^0}=133.5$\,GeV, and $m_{\tilde{\chi}_1^0}=44.6$\,GeV (benchmark 2 from Table\,\ref{tab:benchmarks} in Supplemental Material)
where $f$ is an SM fermion, considering 11 possible SM backgrounds for this process.
The \texttt{XGBOOST} model, trained with 21 kinematic variables, is used to discriminate the signal benchmark from each background class by computing the significance of observing the signal over the background events. At the $\sqrt{s}=14$\,TeV LHC with 137\,fb$^{-1}$ of integrated luminosity~($\mathcal{L}$), we expect to observe 763 signal events and 987 background events for a threshold of 0.9 on our \texttt{XGBOOST} output.
Adding a 20\%~(50\%) systematic uncertainty translates to a significance (using the formula in Ref.\,\cite{Adhikary:2020cli}) of 3.1~(1.3). We present our results for $\sqrt{s}=14$\,TeV to make it easier to translate to the case of Run-3 ($\sqrt{s}=13.6$\,TeV) and HL-LHC ($\sqrt{s}=14$\,TeV) as the cross-sections are not expected to change much.
We find that the result sensitively depends on the systematic uncertainty, which can 
have a significant impact for light electroweakinos. 

In summary, this letter shows that the current experiments, especially the recent results from electroweakino searches at the LHC and dark matter DD measurements at the LZ, have severely constrained the $\mu>0$ scenario, with the $Z$-funnel being completely excluded, and only very heavy higgsinos allowed in the $h$-funnel.
For the $\mu<0$ scenario, the allowed parameter space consists of either higgsinos heavier than $\sim850$\,GeV in the $h$-funnel or restricted to a narrow region of light higgsinos having mass of 120-155\,GeV in the $Z$ and $h$-funnels, for a light neutralino thermal DM in the pMSSM with 10 free parameters. 
Light right-handed staus, still allowed by the LHC, can have a mild affect on the relic density of the lightest neutralino, which 
is investigated in detail in our future work (Ref.\,\cite{future-work}).
Moreover, presence of light staus can affect the collider constraints on higgsinos, in regions of the parameter space where the latter decay into the former with significant branching fractions.

The current status of light higgsinos in the mass range of 125-150\,GeV from the present electroweakino constraints is not completely clear in the pMSSM, since the experimental results are presented for simplified models.
Therefore, experimental collaborations must zoom in on this region and provide definitive answers, covering all possibilities of light higgsinos (including multiple decay modes), which can be an interesting target for Run-3 of LHC.
Such light higgsinos are also motivated by naturalness arguments.
To conclude, at present, we are at a very exciting juncture where the experiments lined up in the near future might exclude the possibility of a light neutralino thermal DM in the MSSM altogether, or we might be very close to starting to observe the first direct hints of new physics at the LHC.

We thank Sabine Kraml for the useful discussion and help related to the \texttt{SModelS} package. The work of GB and RMG was funded in part by the Indo-French Centre for the Promotion of Advanced Research, Grant no: 6304-2. RMG wishes to acknowledge the support of Indian National Science Academy under the award of INSA Senior Scientist.
B.B. and R.S. thank Prabhat Solanki and Camellia Bose for useful discussions. R.K.B. thanks the U.S. Department of Energy for the financial support under grant number DE-SC0016013. 

\bibliographystyle{utphys}

\begin{thebibliography}{10}
\bibitem{footnote2}
The input parameters are read and written in the SLHA file as on-shell
  parameters by the \texttt{FeynHiggs} code. We find no significant changes in
  our result when we use a different spectrum generator, like
  \texttt{SoftSUSY-4.1.17}, which provides the output SLHA with $\overline{DR}$
  parameters.
\bibitem{footnote3}
Non-standard cosmological scenario can impact the results of the current study.
  We are performing a detailed study of the possibility for a non-standard
  cosmological scenario in a future work\,\cite{future-work}.
\bibitem{footnote4}
Note that the contribution of $h/H$ bosons to the SI DD cross-section involves
  the product of couplings $g_{h\tilde{\chi}_1^0\tilde{\chi}_1^0}\times
  g_{hqq}$ and $g_{H\tilde{\chi}_1^0\tilde{\chi}_1^0}\times g_{Hqq}$, where $q$
  can be both up/down-type quarks. The magnitudes and signs of
  $g_{h\tilde{\chi}_1^0\tilde{\chi}_1^0}$,
  $g_{H\tilde{\chi}_1^0\tilde{\chi}_1^0}$ couplings depend on $\mu$ and
  tan\,$\beta$. The $g_{Huu}$ and $g_{Hdd}$ couplings depend on tan\,$\beta$
  with a relative sign difference between them.
\bibitem{footnote}
A smaller $R$-value indicates that the parameter space point lies way outside
  the current limit, whereas a $R$-value close to 1 indicates that it lies just
  on the border.
\end{thebibliography}

\providecommand{\href}[2]{#2}\begingroup\raggedright\endgroup

\clearpage

\appendix

\onecolumngrid

\setcounter{figure}{0} 
\setcounter{page}{1}

\section*{Supplemental Material}

\subsection*{Analyses sensitive for light higgsinos and benchmarks}

\begin{figure}[hbt!]
    \centering
    \includegraphics[width=0.45\textwidth]{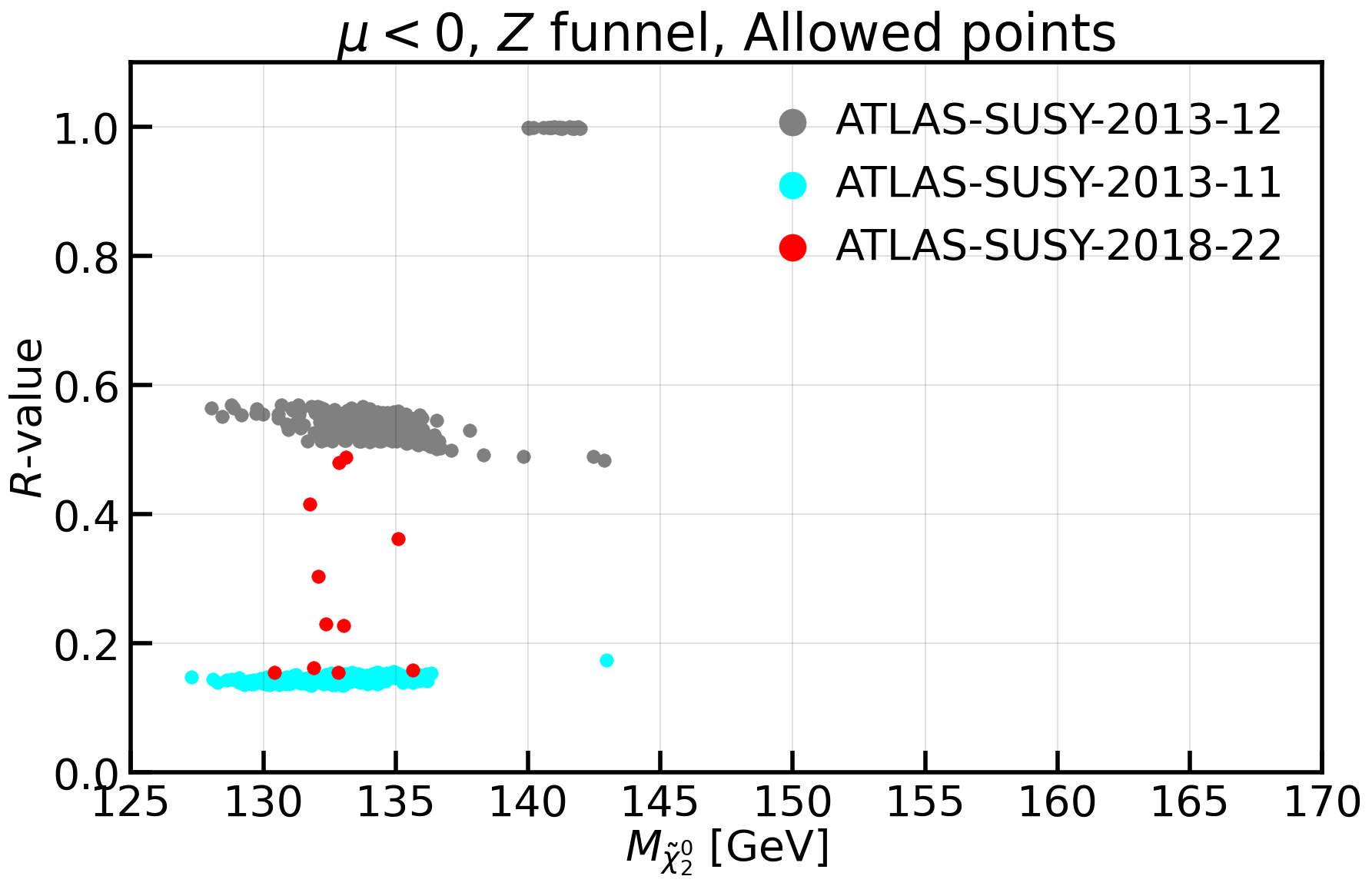}~
    \includegraphics[width=0.45\textwidth]{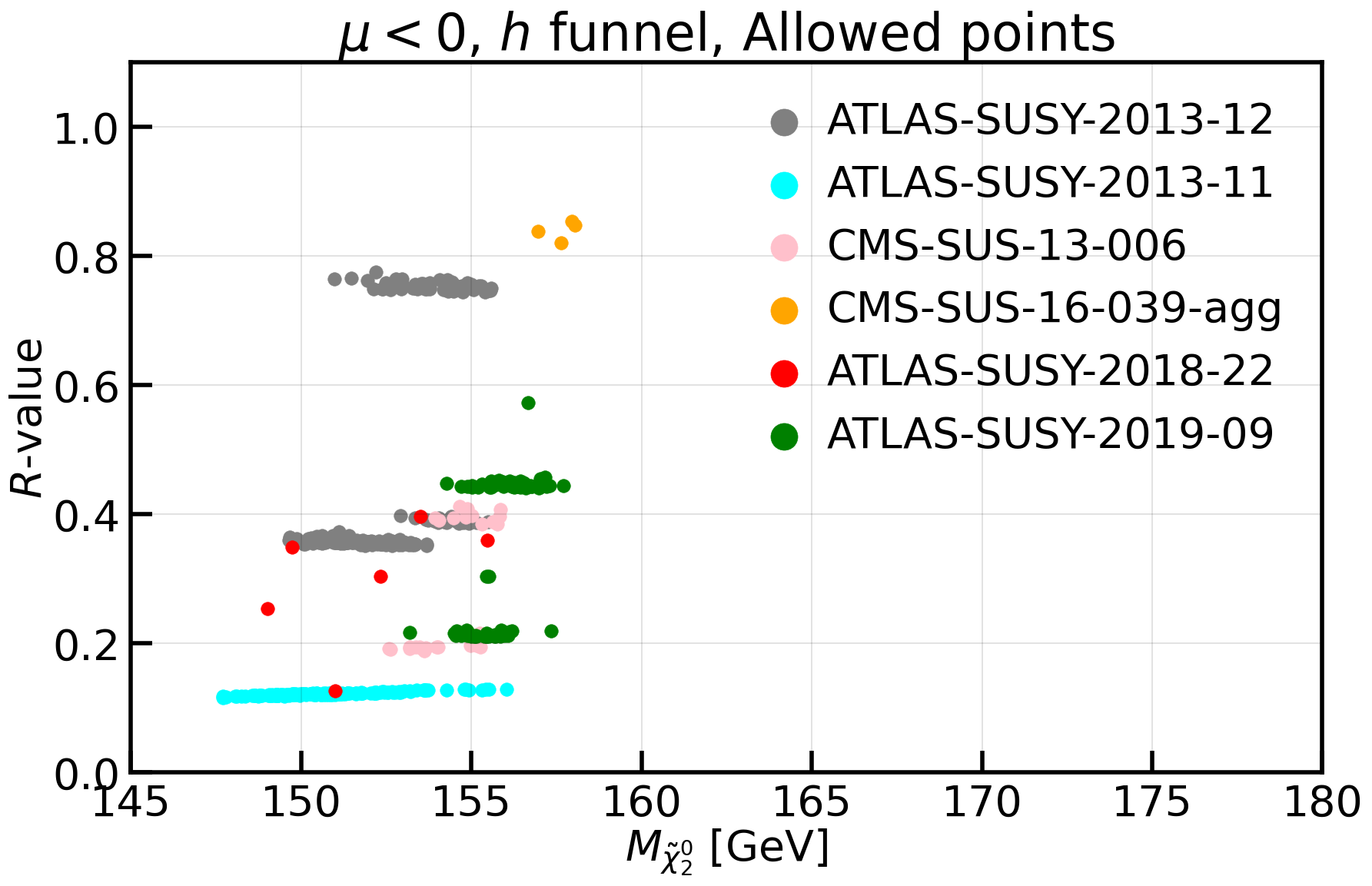}
    \caption{For the lighter higgsinos allowed from \texttt{SModelS}, the range of $M_{\tilde{\chi}_2^0}$ mass for $\mu<0$ in the $Z$ funnel ({\it left}) and in the $h$ funnel ({\it right}) where the different colors depict the most sensitive analysis for each point with the $R$-values of the analysis in the $y$-axis.}
    \label{fig:analysis}
\end{figure}

We have found from our exhaustive scan that for $\mu>0$, we can have allowed points only in the $h$ funnel with $\mu\gtrsim850$\,GeV. For $\mu<0$, we have three allowed regions $-$ very small $\mu$ (125-145\,GeV) in the $Z$ funnel, very small $\mu$ (140-155\,GeV) and large $\mu\gtrsim850$\,GeV in the $h$ funnel. For the lighter higgsinos, we present all the sensitive analyses, as given by the \texttt{SModelS} package along with their $R$-values in Fig.\,\ref{fig:analysis} for the $Z$-funnel ({\it left}) and $h$-funnel ({\it right}) in the negative $\mu$ scenario.  

We present a benchmark point from each of the allowed regions in Table\,\ref{tab:benchmarks} along with the scaled spin-independent DD cross-section for these points. We observe that the points in the $h$ funnel of both positive and negative $\mu$ have DD cross-sections very close to the present LZ bound (at most 12-13\% lower than the current limit) and, therefore, can be probed with only a few more days of LZ data. The benchmark from the $Z$ funnel in $\mu<0$ has a very small SI DD cross-section and can be probed with the LHC electroweakino searches dedicated for light higgsinos.

\begin{table}[hbt!]
    \resizebox{0.95\textwidth}{!}{
    \centering
    \begin{tabular}{c|c|c|c|c}
    \hline
    \multicolumn{3}{c|}{Benchmarks ({\it mass parameters in} GeV)} & $m_{h} [\Delta_{M_h}^{FH}]~$[GeV] & $\sigma_{SI}\times\xi\times10^{-10}$ [pb] \\\hline
    \multirow{2}{*}{$\mu>0$} & \multirow{2}{*}{$h$-funnel} & $M_t=173.21$, $M_1=62.5$, $M_2=2000$, $\mu=1000$, tan$\beta=5$, $M_A=3000$, & \multirow{2}{*}{125.38~[$\pm$0.97]} & \multirow{2}{*}{0.151} \\
    &  & $M_{\tilde{Q}_{3L}}=10000$, $M_{\tilde{t}_{R}}=10000$, $M_{\tilde{b}_{R}}=10000$, $A_t=10000$, $M_3=3000$ & &  \\\hline
    \multirow{6}{*}{$\mu<0$} & \multirow{2}{*}{$Z$-funnel}
                & $M_t=173.21$, $M_1=44$, $M_2=2000$, $\mu=-124$, tan$\beta=5$, $M_A=3000$, & \multirow{2}{*}{125.88~[$\pm$0.96]} & \multirow{2}{*}{$0.746\times10^{-3}$} \\
        &       & $M_{\tilde{Q}_{3L}}=10000$, $M_{\tilde{t}_{R}}=10000$, $M_{\tilde{b}_{R}}=10000$, $A_t=10000$, $M_3=3000$ &  & \\\cline{2-5}
        & \multirow{4}{*}{$h$-funnel} & $M_t=173.21$, $M_1=68$, $M_2=2000$, $\mu=-150$, tan$\beta=50$, $M_A=3000$,& \multirow{2}{*}{125.67~[$\pm$0.63]} & \multirow{2}{*}{0.143} \\
        &    & $M_{\tilde{Q}_{3L}}=5000$, $M_{\tilde{t}_{R}}=5000$, $M_{\tilde{b}_{R}}=5000$, $A_t=-5000$, $M_3=3000$ & & \\\cline{3-5}
        &    & $M_t=173.21$, $M_1=$, $M_2=2000$, $\mu=-1000$, tan$\beta=4.5$, $M_A=3000$, & \multirow{2}{*}{125.15~[$\pm$0.99]} & \multirow{2}{*}{0.150} \\
        &   & $M_{\tilde{Q}_{3L}}=10000$, $M_{\tilde{t}_{R}}=10000$, $M_{\tilde{b}_{R}}=10000$, $A_t=10000$, $M_3=3000$ &  & \\\hline
    \end{tabular}}
    \caption{Parameters corresponding to four benchmark points satisfying all the present constraints from the $\mu>0$ and $\mu<0$ scenarios along with their scaled SI DD cross-sections. The mass of the Higgs boson $M_{h}$ and the uncertainty in $M_h$ computed by FeynHiggs~($\Delta_{M_h}^{FH}$) are also shown.}
    \label{tab:benchmarks}
\end{table}

\subsection*{Details of the \texttt{XGBOOST} analysis}

Here, we discuss the details of our analysis for the light higgsino benchmark.
We study 11 possible SM backgrounds for this process $-$ $lll\nu$ ($l\equiv e,\mu,\tau$), $ZZ$, $t\bar{t}$, $VVV$, $Wh$, $Zh$, ggF and VBF production of $h$ with $h\rightarrow ZZ^*$, $t\bar{t}h$, $t\bar{t}W$, and $t\bar{t}Z$.
We restrict to the leptonic final state which is cleaner for a lighter benchmark, such as ours. We perform an analysis of the $3l+\met$ final state where we require exactly three leptons satisfying $p_T>25,25,20$\,GeV and $|\eta|<2.4$, and we have put a veto on $b$-jets with $p_T>30$\,GeV and $|\eta|<2.5$. In our signal benchmark, since we do not have any on-shell $Z$-boson, we also veto events where the invariant mass of a pair of same flavor opposite sign (SFOS) leptons lie within 10\,GeV window of $m_{Z}=91.2$\,GeV. After these preselections, we train our signal and background samples using \texttt{XGBOOST} with a set of the following variables:

\begin{itemize}
    \item Transverse momenta ($p_T$) of the three leptons
    \item Transverse mass ($M_T$) and contransverse mass ($M_{CT}$) of each of the three leptons with the $\met$
    \item Minimum and maximum values of $\Delta R$ between opposite sign lepton pairs along with their $\Delta\eta$ values
    \item Invariant mass of the opposite sign lepton pairs with minimum and maximum $\Delta R$
    \item Missing transverse momentum
    \item Number of jets in the event with the $p_T$ of the two leading jets
    \item Scalar sum of $p_T$ of all the jets in the event ($H_T$)
    \item Invariant mass of the three leptons
\end{itemize}

Following are the hyperparameters of the XGBOOST model:\\ \texttt{`objective':`multi:softprob', `colsample\_bytree':0.3, `learning\_rate':0.1, `num\_class':12, `max\_depth':7, `alpha':5, `eval\_metric':`mlogloss', `num\_round':1000, `early\_stopping\_rounds':3}

\subsection*{Prospects at Future Lepton Colliders}

The future lepton colliders like ILC and CEPC will be crucial for precision measurements of Higgs boson. The projected upper limit on the invisible branching of the Higgs boson is 0.4\% at ILC\,\cite{Asner:2013psa} and 0.3\% at CEPC\,\cite{An:2018dwb}. Although these can probe a significant part of the allowed parameter space in the $\mu<0$ case, we still have regions with Br($h\rightarrow$invisible$<0.003$) in both the $Z$ and $h$-funnels. 
In the $\mu<0$ case, the partial decay width of the $Z$ boson to $\tilde{\chi}_1^0$ ($\Gamma_{\rm inv}^{\rm new}$) is always less than 0.1\,MeV for the allowed parameter region that we obtain. Therefore, we do not expect the Giga-$Z$ option of ILC, which is expected to have a modest improvement over LEP\,\cite{Carena:2003aj}, to be sensitive to this region.

\end{document}